\documentstyle[12pt]{article}
\begin{document}

\title{Constraints on Vacuum Oscillations\\ from Recent Solar Neutrino Data}

\author{C. E. C. Lima$^{1,}$\thanks{ceclima@cbpf.br} and H. M. Portella$^{2,}$\thanks{portella@cbpf.br}}

\maketitle

\begin{center}

$^1${\it Centro Brasileiro de Pesquisas F\'{\i}sicas, Rio de Janeiro, RJ, 22290-180, Brazil}

$^2${\it Universidade Federal Fluminense, Niter\'oi, RJ, 24210-340, Brazil}

\vspace{.8 true cm}

\abstract{A detailed study of the solar neutrino vacuum oscillation is made taking into account three neutrino flavours and seasonal effect. A set of $\delta m_{21}^2$ x $\sin^2(2\omega)$ regions is calculated for a range of the parameter $\sin^2(2\phi)$ from 0 to 1, with and without the inclusion of the recoil-electron spectrum in the rates. The averaged survival probabilities for $\nu_e$ as a function of the energy are obtained, what reveal that solutions with values of $\sin^2(2\phi) \geq 0.50$ and the maximum exclusion of $^7Be$ neutrinos and minimum exclusion of $pp$ neutrinos give a better explanation for the suppression rates of all detectors.} 

\end{center}

\newpage

\section{Introduction}

\noindent

For three decades the deficit on the observed flux of solar neutrinos compared to that one expected from Standard Solar Models (SSM) has puzzled the scientists and originated several explanations for this problem. These solutions are divided into astrophysical and beyond the standard model \cite{Bahcall-1}. Nowadays, the astrophysical solutions have been highly disfavoured by the helioseismological data which are well explained by standard solar models, however, solutions with new physics, in special the neutrino oscillations, that can be in vacuum \cite{Pontecorvo} or in matter (by MSW effect) \cite{MSW}, gives a good fit to the solar neutrino data.

In the present work we will focus attention only on vacuum oscillations, which occur in the neutrino way from the Sun to Earth. The evidence of the existence of vacuum oscillations come from the observation of the seasonal effect. Because the Earth orbit is not circular, the variations in the Earth-Sun distance produce a difference in the oscillation probability what can be very important for neutrino line sources, in special, the $^7Be$ (0.861 MeV) and is to be confirmed by the Borexino detector \cite{Borexino}. In our study we will deal with two and three neutrino flavours to try to explain the solar neutrino problem. We will vary the parameter $\sin^2(2\phi)$ that is concerned with the $\nu_e {^{\rightarrow}_\leftarrow} \nu_\tau$ oscillations in order to discover which scenario (two or three generations) gives a better description for the experimental data. This analysis will be made by the calculation of the 
$\delta m_{21}^2$ x $\sin^2(2\omega)$ regions considering first only the rates of the detectors without the spectral distortion, caused by the recoil-electron spectrum in Kamiokande and Super-Kamiokande, and after including these spectrum within the calculation. An investigation in the $\nu_e$ survival probability will be made, what will reveal the behavior of the neutrino as a function of the energy.

In section 2 we give the relations needed to calculate the mass regions and mixing angles allowed
by the experiments. In section 3 we present the $\nu_e$ survival probability as well the $\nu_\mu$ and $\nu_\tau$ conversion probabilities. In section 4 we analise the obtained data to describe the probabilities as a function of the energy for a set of values of $\sin^2(2\phi)$. In section 5 we present the conclusions of this work. 

\section{Suppression Rates}

\noindent

The first procedure is to calculate the suppression rates of the experiments using neutrino oscillations and compare with the observed ones to search for the values of the parameters $\delta m_{21}^2$ and $\sin^2(2\omega)$ that fit the five experiments (Homestake \cite{Homestake}, GALLEX \cite{GALLEX}, SAGE \cite{SAGE}, Kamiokande \cite{Kamiokande} and Super-Kamiokande \cite{Super-K}) together. The suppression rate, to the experimental case, is obtained by dividing 
the capture rate of the detector to that one expected from the SSM and theoretically, dividing the capture rate calculated using neutrino oscillations to the SSM one. We will make this calculation using the data from the BP98 Standard Solar Model \cite{BP98} with 99\% C.L.

The Solar Neutrino Problem is evidenced when we compare the theoretical capture rates with those obtained experimentally. When this comparison is made we discover that these rates never match, giving values varying from 33\% for Homestake to 60\% for GALLEX. See table 1.

\begin{table}
\caption{Solar neutrino data and BP98 SSM predictions [10]. The second and third columns show the measured and calculated capture rates for each detector. The values are given in SNU except for Kamiokande and Super-Kamiokande, where the parameter is the $^8B$ neutrino flux, in units of $10^6 cm^{-2} s^{-1}$. The suppression rates are set on the fourth column for 1 $\sigma$ errors.}
\begin{center}
\begin{tabular}{|l|l|l|l|} \hline
Experiment & Result & BP98 SSM & Result/SSM \\ \hline
Homestake \cite{Homestake} & $2.56 \pm 0.16 \pm 0.16$ & $7.7^{+1.2}_{-1.0}$ & $0.33 \pm 0.03$ \\
GALLEX \cite{GALLEX} & $77.5 \pm 6.2 ^{+4.3}_{-4.7}$ & $129^{+8.0}_{-6.0}$ & $0.60 \pm 0.06$ \\ 
SAGE \cite{SAGE} & $66.6 ^{+7.8}_{-8.1}$ & $129^{+8.0}_{-6.0}$ & $0.52 \pm 0.06$ \\
Kamiokande \cite{Kamiokande} & $2.80\pm 0.19\pm 0.33$ & $5.15^{+1.0}_{-0.7}$ & $0.54\pm 0.07$ \\
Super-Kamiokande \cite{Super-K} & $2.44 \pm 0.05 ^{+0.09}_{-0.07}$ & $5.15^{+1.0}_{-0.7}$ & $0.47\pm 0.02$ \\ \hline
\end{tabular}
\end{center}
\end{table}

The solar neutrino suppression rate using neutrino oscillation is given by 

\begin{equation}
R = \frac{S_{\mbox{\tiny OSC}}}{S_{\mbox{\tiny SSM}}} 
\label{rate}
\end{equation}

\noindent
where

\begin{equation}
S_{\mbox{\tiny OSC}} = \sum_i \Phi_i \int_{{\mbox{\tiny E}}_{\mbox{\tiny  min}}}^{{\mbox{\tiny E}}_{\mbox{\tiny max}}} \eta_i{\mbox{\tiny (E}}_\nu\mbox{\tiny )} \sigma{\mbox{\tiny (E}}_\nu\mbox{\tiny )} P_{\nu_e \nu_e}{\mbox{\tiny (E}}_\nu\mbox{\tiny )} dE_\nu
\label{S-OSC1}
\end{equation}

\noindent
and $S_{\mbox{\tiny SSM}}$ is the same expression with $P_{\nu_e \nu_e}{\mbox{\tiny (E}}_\nu\mbox{\tiny )} = 1$

The equation (\ref{S-OSC1}) is available for Homestake, GALLEX and SAGE experiments, where $i$ runs for each neutrino source, $\Phi_i$ is the total neutrino flux from the source $i$, $\eta_i{\mbox{\tiny (E}}_\nu\mbox{\tiny )}$ is its normalized neutrino energy spectrum, $\sigma{\mbox{\tiny (E}}_\nu\mbox{\tiny )}$ is the cross section for the experiment considered and $P_{\nu_e \nu_e}{\mbox{\tiny (E}}_\nu\mbox{\tiny )}$ is the electron neutrino survival probability. The neutrino sources that we consider for these experiments are $pp$, $^7Be$, $^8B$, $pep$, $^{13}N$ and $^{15}O$, except for Homestake that is not sensitive to $pp$ neutrinos.

For Super-Kamiokande we consider the $^8B$ neutrino source only and we have to take into account the fact that this experiment is also sensitive to the $\nu_\mu,e$ and $\nu_\tau,e$ scattering, so $S_{\mbox{\tiny OSC}}$ becomes

\begin{equation}
S_{\mbox{\tiny OSC}} = \Phi \int_{{\mbox{\tiny E}}_{\mbox{\tiny  min}}}^{{\mbox{\tiny E}}_{\mbox{\tiny max}}} \eta{\mbox{\tiny (E}}_\nu\mbox{\tiny )} \left\{ P_{\nu_e \nu_e} {\mbox{\tiny (E}}_\nu\mbox{\tiny )}\sigma_{\nu_e}{\mbox{\tiny (E}}_\nu\mbox{\tiny )} + P_{\nu_\mu \nu_e}{\mbox{\tiny (E}}_\nu\mbox{\tiny )} \sigma_{\nu_\mu}{\mbox{\tiny (E}}_\nu\mbox{\tiny )} + P_{\nu_\tau \nu_e}{\mbox{\tiny (E}}_\nu\mbox{\tiny )} \sigma_{\nu_\tau}{\mbox{\tiny (E}}_\nu\mbox{\tiny )} \right\} dE_\nu 
\label{S-OSC1b}
\end{equation}

\noindent
since $\sigma_{\nu_\mu}{\mbox{\tiny (E}}_\nu\mbox{\tiny )} = \sigma_{\nu_\tau}{\mbox{\tiny (E}}_\nu\mbox{\tiny )}$ and $P_{\nu_\mu \nu_e}{\mbox{\tiny (E}}_\nu\mbox{\tiny )} + P_{\nu_\tau \nu_e}{\mbox{\tiny (E}}_\nu\mbox{\tiny )} = 1 - P_{\nu_e \nu_e}{\mbox{\tiny (E}}_\nu\mbox{\tiny )}$ the equation (\ref{S-OSC1b}) is simplified to

\begin{equation}
S_{\mbox{\tiny OSC}} = \Phi \int_{{\mbox{\tiny E}}_{\mbox{\tiny  min}}}^{{\mbox{\tiny E}}_{\mbox{\tiny max}}} \eta{\mbox{\tiny (E}}_\nu\mbox{\tiny )} \left\{ P_{\nu_e \nu_e} {\mbox{\tiny (E}}_\nu\mbox{\tiny )}\sigma_{\nu_e}{\mbox{\tiny (E}}_\nu\mbox{\tiny )} + (1 - P_{\nu_e \nu_e}{\mbox{\tiny (E}}_\nu\mbox{\tiny )}) \sigma_{\nu_\mu}{\mbox{\tiny (E}}_\nu\mbox{\tiny )} \right\} dE_\nu
\label{S-OSC1c}
\end{equation}

For the Kamiokande experiment only the $^8B$ source is considered again but the sensitivity to $\nu_\tau$ is neglected.

As Super-Kamiokande and Kamiokande are neutrino-electron scattering detectors, different from the
other three which use neutrino absorption, we can also take into account the energy resolution in order to observe the influence of the spectral distortion caused by the recoil-electrons in the suppression rates, giving the following expressions for $S_{\mbox{\tiny OSC}}$ and $S_{\mbox{\tiny SSM}}$

$$S_{\mbox{\tiny OSC}} = N_{\mbox{\tiny 0}} \Phi \int_{{\mbox{\tiny E}}_{\mbox{\tiny  min}}}^{{\mbox{\tiny E}}_{\mbox{\tiny max}}} dE_\nu \eta{\mbox{\tiny (E}}_\nu\mbox{\tiny )}\int_{{\mbox{\tiny T}}_{\mbox{\tiny min}}-{\mbox{\tiny m}}_{\mbox{\tiny e}}}^{{\mbox{\tiny T}}_{\mbox{\tiny max}}-{\mbox{\tiny m}}_{\mbox{\tiny e}}}dT\int_{\mbox{\tiny 0}}^{{\mbox{\tiny T'}}_{\mbox{\tiny max}}}dT' R\mbox{\tiny (T',T)}\left\{ P_{\nu_e \nu_e} {\mbox{\tiny (E}}_\nu\mbox{\tiny )}\frac{d\sigma_{\nu_e}{\mbox{\tiny (T',E}}_\nu\mbox{\tiny )}}{dT'} + \right.$$

\begin{equation}
\left. + P_{\nu_\mu \nu_e}{\mbox{\tiny (E}}_\nu\mbox{\tiny )} {\frac{d\sigma_{\nu_\mu}{\mbox{\tiny (T',E}}_\nu\mbox{\tiny )}}{dT'}}+P_{\nu_\tau \nu_e}{\mbox{\tiny (E}}_\nu\mbox{\tiny )} {\frac{d\sigma_{\nu_\tau}{\mbox{\tiny (T',E}}_\nu\mbox{\tiny )}}{dT'}} \right\}
\label{S-OSC2}
\end{equation}

\begin{equation}
S_{\mbox{\tiny SSM}} = N_{\mbox{\tiny 0}} \Phi \int_{{\mbox{\tiny E}}_{\mbox{\tiny  min}}}^{{\mbox{\tiny E}}_{\mbox{\tiny max}}} dE_\nu \eta{\mbox{\tiny (E}}_\nu\mbox{\tiny )}\int_{{\mbox{\tiny T}}_{\mbox{\tiny min}}-{\mbox{\tiny m}}_{\mbox{\tiny e}}}^{{\mbox{\tiny T}}_{\mbox{\tiny max}}-{\mbox{\tiny m}}_{\mbox{\tiny e}}}dT \int_{\mbox{\tiny 0}}^{{\mbox{\tiny T'}}_{\mbox{\tiny max }}}dT' R\mbox{\tiny (T',T)} {\frac{d\sigma_{\nu_e}{\mbox{\tiny (T',E}}_\nu\mbox{\tiny )}}{dT'}}
\label{S-SSM2}
\end{equation}

The same procedure used to simplify the eq.(\ref{S-OSC1b}) is taken here, what leaves eq.(\ref{S-OSC2}) as 

$$S_{\mbox{\tiny OSC}} = N_{\mbox{\tiny 0}} \Phi \int_{{\mbox{\tiny E}}_{\mbox{\tiny  min}}}^{{\mbox{\tiny E}}_{\mbox{\tiny max}}} dE_\nu \eta{\mbox{\tiny (E}}_\nu\mbox{\tiny )}\int_{{\mbox{\tiny T}}_{\mbox{\tiny min}}-{\mbox{\tiny m}}_{\mbox{\tiny e}}}^{{\mbox{\tiny T}}_{\mbox{\tiny max}}-{\mbox{\tiny m}}_{\mbox{\tiny e}}}dT \int_{\mbox{\tiny 0}}^{{\mbox{\tiny T'}}_{\mbox{\tiny max}}}dT' R\mbox{\tiny (T',T)}\left\{ P_{\nu_e \nu_e}{\mbox{\tiny (E}}_\nu\mbox{\tiny )} {\frac{d\sigma_{\nu_e}{\mbox{\tiny (T',E}}_\nu\mbox{\tiny )}}{dT'}} + \right.$$

\begin{equation}
\left.(1 - P_{\nu_e \nu_e}{\mbox{\tiny (E}}_\nu\mbox{\tiny )}) {\frac{d\sigma_{\nu_\mu}{\mbox{\tiny (T',E}}_\nu\mbox{\tiny )}}{dT'}} \right\}
\label{S-OSC3}
\end{equation}

The energy resolution is described by

\begin{equation}
R\mbox{\tiny (T',T)} = \frac{1}{\Delta_{T'}\sqrt{2\pi}} exp\left\{- \frac{(T' - T + \delta )^2}{2 \Delta^2_{T'}}\right\}
\end{equation}

\noindent
where $T$ and $T'$ are the measured and true electron kinetic energy, respectively, and the limit $T'_{\mbox{\tiny max}}$ in the third integral of the eq.(\ref{S-OSC3}) is the maximum kinetic energy that an electron can achieve given the neutrino energy $E_\nu$, 
$T'_{\mbox{\tiny max}}~=~E_\nu /(1~+~m_e/2 E_\nu)$. For more details see \cite{Bahcall-87,  Bahcall-1}.   

The limits of energy used to calculate these rates are not the same the neutrinos have in the Sun, so we have to calculate the normalized detector sensitivity to know these limits, as follows

\begin{equation}
{\mbox{\it f}}_i {\mbox{\tiny (E}}_\nu\mbox{\tiny )} = \frac{\sigma {\mbox{\tiny (E}}_\nu\mbox{\tiny )} \eta_i {\mbox{\tiny (E}}_\nu\mbox{\tiny )}}{\int_E \sigma {\mbox{\tiny (E}}_\nu\mbox{\tiny )} \eta_i {\mbox{\tiny (E}}_\nu\mbox{\tiny )} dE_\nu }
\end{equation}

We can notice in the figure 1 the existence of regions where the spectral sensitivity is zero.

\section{Electron Neutrino Survival Probability}

\noindent

For three neutrino flavours the $\nu_e$ survival probability is written as \cite{Fogli}

\begin{equation}
P_{\nu_e \nu_e}{\mbox{\tiny (E}}_\nu\mbox{\tiny )} = 1 - C^4_\phi S^2_{2\omega} \sin^2(\frac{\delta m_{21}^2 x}{4 \hbar c E}) - C^2_\omega S^2_{2\phi} \sin^2(\frac{\delta m_{31}^2 x}{4 \hbar c E})
- S^2_\omega S^2_{2\phi} \sin^2(\frac{\delta m_{32}^2 x}{4 \hbar c E})
\label{Pee}
\end{equation}

\noindent
where the terms $C^4_\phi$, $S^2_{2\omega}$, $S^2_{\omega}$, $C^2_{\omega}$ and $S^2_{2\phi}$ are short forms for $\cos^4(\phi)$, $\sin^2(2\omega)$, $\sin^2(\omega)$, $\cos^2(\omega)$ and $\sin^2(2\phi)$, respectively. 

In this work we will consider the following mass hierarchy \cite{Barger}

\begin{equation}
m1 << m2 << m3
\end{equation}

\noindent
what leads us to $\delta m_{31}^2 >> \delta m_{21}^2$ and $\delta m_{32}^2 \approx \delta m_{31}^2$, since $\delta m_{32}^2 = \delta m_{31}^2 - \delta m_{21}^2$, thus we can simplify the equation (\ref{Pee}) to the form

\begin{equation}
P_{\nu_e \nu_e}{\mbox{\tiny (E}}_\nu\mbox{\tiny )} = 1 - C^4_\phi S^2_{2\omega} \sin^2(\frac{\delta m_{21}^2 x}{4 \hbar c E}) - S^2_{2\phi} \sin^2(\frac{\delta m_{31}^2 x}{4 \hbar c E})
\label{Pee1}
\end{equation}

Now we have two choices

\noindent
i) if $\delta m_{21}^2 \leq 10^{-11} eV^2$ then the term $\sin^2(\frac{\delta m_{21}^2 x}{4 \hbar c E})$ tends to be very small and we fall in the case of two generations with $\phi$ as the mixing angle and $\delta m_{31}^2$ as the mass parameter; 

\noindent  
ii) if $\delta m_{21}^2 \approx 10^{-10} eV^2$, for the range of energy considered the term $\sin^2(\frac{\delta m_{31}^2 x}{4 \hbar c E})$ can be averaged to $1/2$, giving the final equation for the $\nu_e$ survival probability, which applies for two and three neutrino oscillations

\begin{equation}
P_{\nu_e \nu_e}{\mbox{\tiny (E}}_\nu\mbox{\tiny )} = 1 - \frac{1}{2} S^2_{2\phi} - C^4_\phi S^2_{2\omega} \sin^2(\frac{\delta m_{21}^2 x}{4 \hbar c E})
\label{Pee2}  
\end{equation}

The introduction of the seasonal effect causes a modification in the probability since the distance $x$ varies with the time on the following way \cite{Krastev}

\begin{equation}
x\mbox{\tiny (t)} = x_{\mbox{\tiny 0}} (1 - \epsilon \mbox{\hspace{.1 true cm}} \cos(\frac{2 \pi t}{{\mbox{\Large $\tau$}}}))
\end{equation}

\noindent
where $x_{\mbox{\tiny 0}}$ is the mean Earth-Sun distance ($\approx 1.49$ . $10^{11} m$), $\epsilon$ is the elipticity of the orbit and {\Large $\tau$}~=~365 days. Once we are taking a temporal average over the probability the eq.(\ref{Pee2}) turns. 

\begin{equation}
\label{Peet}
P_{\nu_e \nu_e}{\mbox{\tiny (t,E}}_\nu\mbox{\tiny )} = 1 - \frac{1}{2} S^2_{2\phi} - C^4_\phi S^2_{2\omega} \frac{1}{(t_2 - t_1)} \int_{{\mbox{\tiny t}}_{\mbox{\tiny 1}}}^{{\mbox{\tiny t}}_{\mbox{\tiny 2}}} \sin^2 \left\{\frac{\delta m_{21}^2 x_{\mbox{\tiny 0}}(1 - \epsilon \mbox{\hspace{.1 true cm}} \cos(\frac{2 \pi t}{{\mbox{\large $\tau$}}}))}{4 \hbar c E}\right\} dt  
\end{equation}

\noindent
where we integrate in a time interval $[t_1,t_2]$, which in this work we consider as one year.

For $P_{\nu_\mu \nu_e}{\mbox{\tiny (t,E}}_\nu\mbox{\tiny )}$ and $P_{\nu_\tau \nu_e}{\mbox{\tiny (t,E}}_\nu\mbox{\tiny )}$ we have

\begin{equation}
\label{Pemt}
P_{\nu_\mu \nu_e}{\mbox{\tiny (t,E}}_\nu\mbox{\tiny )} = C^2_\phi S^2_{2\omega} \frac{1}{(t_2 - t_1)} \int_{{\mbox{\tiny t}}_{\mbox{\tiny 1}}}^{{\mbox{\tiny t}}_{\mbox{\tiny 2}}} \sin^2 \left\{\frac{\delta m_{21}^2 x_{\mbox{\tiny 0}}(1 - \epsilon \mbox{\hspace{.1 true cm}} \cos(\frac{2 \pi t}{{\mbox{\large $\tau$}}}))}{4 \hbar c E}\right\} dt  
\end{equation}

\noindent
and

\begin{equation}
\label{Pett}
P_{\nu_\tau \nu_e}{\mbox{\tiny (t,E}}_\nu\mbox{\tiny )} = \frac{1}{2} S^2_{2\phi} - \frac{1}{4} S^2_{2\phi} S^2_{2\omega} \frac{1}{(t_2 - t_1)} \int_{{\mbox{\tiny t}}_{\mbox{\tiny 1}}}^{{\mbox{\tiny t}}_{\mbox{\tiny 2}}} \sin^2 \left\{\frac{\delta m_{21}^2 x_{\mbox{\tiny 0}}(1 - \epsilon \mbox{\hspace{.1 true cm}} \cos(\frac{2 \pi t}{{\mbox{\large $\tau$}}}))}{4 \hbar c E}\right\} dt 
\end{equation}

In figure 2 it is shown the expected maximal and minimal variation of these probabilities as a function of $\sin^2(2\phi)$.

\newpage

\section{Analysis of the Results}

\noindent

The figure 3 illustrates the plots of $\delta m_{21}^2$ x $\sin^2(2\omega)$ considering the seasonal effect. We can notice an increase in the regions that explain the five experiments together as $\sin^2(2\phi)$ runs from 0 to 1. The first column shows the regions without taking into account the recoil-electron spectrum of Kamiokande and Super-Kamiokande and the second column gives the regions with the inclusion of that spectrum, what causes a decrease in the mass regions in the order of 33\% for $\sin^2(2\phi) = 0$ to 23\% for $\sin^2(2\phi) = 1$. The $\chi^2$ analysis was made following the references \cite{Krastev,Calabresu} and is shown in table 2 and we can observe that the inclusion of the recoil-electron spectrum gives a better fit. 

With the $\delta m_{21}^2$ and $\sin^2(2\omega)$ points from figure 3 we can obtain the averaged electron neutrino survival probabilities as a function of the energy, what are ploted on figure 4. In a first sight we observe a considerable decrease of the probability on the energy region dominated by the $^8B$ neutrino and a slight decrease in the $^7Be (0.861 MeV)$ line source, because it lies in a region where there are CNO neutrinos also. As the parameter $\sin^2(2\phi)$ grows, the probability range decrease with the energy. For $\sin^2(2\phi) = 0$, $0.30~\leq~P_{\nu_e \nu_e}~\leq~0.80$ and for $\sin^2(2\phi) = 1$, $0.39~\leq~P_{\nu_e \nu_e}~\leq~0.42$.

To make a better analysis it is useful to adopt the procedure given by Barger {\it et al.}, \cite{Barger}, where the suppression rates of the detectors are given by means of the averaged probabilities. So we have for the detectors the following expressions,

\begin{table}
\caption{Best fit points for the $\delta m_{21}^2$ x $\sin^2(2\omega)$ plots shown on figure 3. Column~1 is for the seasonal effect with the rates only and column~2 is with the inclusion of the recoil-electron spectrum.}
\begin{center}
\begin{tabular}{|l|l|l|l|} \hline
$\sin^2(2\phi)$ & $\chi^2_{min}$ & $\sin^2(2\omega)$ & $\delta m_{21}^2 (eV^2)$ \\ \hline
\multicolumn{4}{|c|}{Column 1} \\ \hline
0    & $9.29$ & $0.76$ & $6.22$ . $10^{-11}$ \\ 
0.25 & $9.95$ & $0.85$ & $3.17$ . $10^{-10}$ \\ 
0.50 & $10.3$ & $0.79$ & $3.25$ . $10^{-10}$ \\ 
0.75 & $10.9$ & $0.91$ & $3.50$ . $10^{-10}$ \\ 
1    & $12.9$ & $0.81$ & $3.70$ . $10^{-10}$ \\ \hline
\multicolumn{4}{|c|}{Column 2} \\ \hline
0    & $4.42$ & $0.79$ & $2.70$ . $10^{-10}$ \\ 
0.25 & $4.78$ & $0.72$ & $2.70$ . $10^{-10}$ \\ 
0.50 & $5.17$ & $0.96$ & $7.88$ . $10^{-10}$ \\ 
0.75 & $5.49$ & $0.79$ & $8.15$ . $10^{-10}$ \\ 
1    & $6.49$ & $0.63$ & $8.15$ . $10^{-10}$ \\ \hline
\end{tabular}
\end{center}
\end{table}

\begin{equation}
\begin{array}{rcl}
R_{Hom} & = & 0.76 \langle P_{\nu_e \nu_e} \rangle_{^8B} + 0.15 \langle P_{\nu_e \nu_e} \rangle_{^7Be} + 0.09 \langle P_{\nu_e \nu_e} \rangle_{CNO,pep} \\
R_{Gal} & = & 0.10 \langle P_{\nu_e \nu_e} \rangle_{^8B} + 0.27 \langle P_{\nu_e \nu_e} \rangle_{^7Be} + 0.09 \langle P_{\nu_e \nu_e} \rangle_{CNO,pep} + 0.54 \langle P_{\nu_e \nu_e} \rangle_{pp} \\
R_{Kam} & = & \langle P_{\nu_e \nu_e} \rangle_{^8B} + \frac{1}{7}\langle P_{\nu_\mu \nu_e} \rangle_{^8B} \\
R_{S-K} & = & \frac{1}{7} + \frac{6}{7}\langle P_{\nu_e \nu_e} \rangle_{^8B}
\end{array}
\end{equation}

\noindent
where for simplification on the calculus we take into account the fact that $\sigma_{\nu_e,e}{\mbox{\tiny (E}}_\nu\mbox{\tiny )} \approx 7 \sigma_{\nu_\mu,e}{\mbox{\tiny (E}}_\nu\mbox{\tiny )}$ in Kamiokande and Super-Kamiokande. In table 3 there are the values (99\% C.L.) of the suppression rates using the probabilities from figure 4 considering the following range of energies: $E_\nu \leq 0.420 MeV$ for $pp$ and $E_\nu = 0.861 MeV$ for $^7Be$. For $^8B$ we used $E_\nu \geq 5.0 MeV$ for the radiochemical detectors and $E_\nu \geq 6.5$ and $7.5 MeV$ for Super-Kamiokande and Kamiokande, respectively. For $CNO$ neutrinos the range is $0.814 \leq E_\nu \leq 1.732 MeV$ for Homestake and $0.233 \leq E_\nu \leq 1.732 MeV$ for the gallium detectors. 

\begin{table}
\caption{Suppression rates for the experiments using the averaged solar neutrino survival probabilities, with 99\% C.L.}
\begin{center}
\begin{tabular}{|l|l|l|l|l|} \hline
$\sin^2(2\phi)$ & Homestake & Gallium & Kamiokande & Super-Kam \\ \hline
\multicolumn{5}{|c|}{rates only} \\ \hline
0    & $0.52 \pm 0.19$ & $0.56 \pm 0.08$ & $0.49 \pm 0.10$ & $0.50 \pm 0.11$ \\ 
0.25 & $0.45 \pm 0.11$ & $0.52 \pm 0.04$ & $0.46 \pm 0.11$ & $0.49 \pm 0.09$ \\  
0.50 & $0.42 \pm 0.07$ & $0.47 \pm 0.03$ & $0.45 \pm 0.09$ & $0.48 \pm 0.07$ \\  
0.75 & $0.41 \pm 0.04$ & $0.43 \pm 0.02$ & $0.44 \pm 0.05$ & $0.48 \pm 0.04$ \\  
1    & $0.41 \pm 0.01$ & $0.41 \pm 0.01$ & $0.43 \pm 0.02$ & $0.48 \pm 0.01$ \\ \hline
\multicolumn{5}{|c|}{rates + recoil-electron spectrum} \\ \hline
0    & $0.50 \pm 0.17$ & $0.55 \pm 0.08$ & $0.52 \pm 0.13$ & $0.53 \pm 0.14$ \\ 
0.25 & $0.45 \pm 0.12$ & $0.49 \pm 0.06$ & $0.51 \pm 0.15$ & $0.52 \pm 0.12$ \\  
0.50 & $0.43 \pm 0.09$ & $0.45 \pm 0.03$ & $0.47 \pm 0.13$ & $0.50 \pm 0.10$ \\  
0.75 & $0.43 \pm 0.06$ & $0.42 \pm 0.02$ & $0.44 \pm 0.08$ & $0.50 \pm 0.06$ \\  
1    & $0.43 \pm 0.03$ & $0.42 \pm 0.01$ & $0.44 \pm 0.03$ & $0.50 \pm 0.02$ \\ \hline
\end{tabular}
\end{center}
\end{table}

\begin{table}
\caption{Suppression rates using the averaged solar neutrino survival probabilities, with 99\% C.L., considering maximal suppression of $^7Be$ neutrinos for Homestake and gallium experiments and minimal suppression of $pp$ neutrinos for gallium.}
\begin{center}
\begin{tabular}{|l|l|l|} \hline
$\sin^2(2\phi)$ & Homestake & Gallium \\ \hline
\multicolumn{3}{|c|}{rates only} \\ \hline
0    & $0.44 \pm 0.16$ & $0.67 \pm 0.14$ \\ 
0.25 & $0.38 \pm 0.11$ & $0.60 \pm 0.10$ \\  
0.50 & $0.36 \pm 0.08$ & $0.51 \pm 0.06$ \\  
0.75 & $0.35 \pm 0.04$ & $0.44 \pm 0.02$ \\  
1    & $0.34 \pm 0.02$ & $0.42 \pm 0.01$ \\ \hline
\multicolumn{3}{|c|}{rates + recoil-electron spectrum} \\ \hline
0    & $0.43 \pm 0.17$ & $0.66 \pm 0.14$ \\ 
0.25 & $0.38 \pm 0.12$ & $0.59 \pm 0.11$ \\  
0.50 & $0.36 \pm 0.09$ & $0.49 \pm 0.07$ \\  
0.75 & $0.34 \pm 0.05$ & $0.42 \pm 0.02$ \\  
1    & $0.33 \pm 0.02$ & $0.42 \pm 0.01$ \\ \hline
\end{tabular}
\end{center}
\end{table}

A comparison between table 3 and the experimental data from table 1 shows us that the suppression rates for Homestake are over the experimental ones for all values of $\sin^2(2\phi)$, but as this parameter tends to 1 the fit of the values turns better and the rates of Homestake can be well explained within $3.5\sigma$ for $\sin^2(2\phi) \geq 0.75$, considering the rates with or without recoil-electron spectrum\footnote{Note that Homestake, GALLEX and SAGE are not affected individualy by the recoil spectrum as Kamiokande and Super-Kamiokande are, but we are dealing with the data from the plots that fits the five experiments together.}. This is also verified in Super-Kamiokande for $\sin^2(2\phi) \geq 0.5$. For the gallium experiments all rates are checked for any $\sin^2(2\phi)$ with $3.5\sigma$, but small values of this parameter  give a better agreement, so we can check GALLEX and SAGE within an experimental $2\sigma$ error for $\sin^2(2\phi) \leq 0.25$. For Kamiokande the data are explained for all values of $\sin^2(2\phi)$ within $3.5\sigma$, but for $\sin^2(2\phi) = 1$ it is checked with $2\sigma$.

A better fit for Homestake is obtained considering the maximum possible suppression of $^7Be$ neutrino\footnote{The total suppression of the $^7Be$ neutrino only occurs for $\sin^2(2\phi)~=~0$ (see figure 2).} (this suppression is limited by eq.\ref{Peet}); in special for $\sin^2(2\phi) \geq 0.5$ this agreement is found whitin $3\sigma$ error. Unfortunatelly this fact leads to a decrease in the GALLEX and SAGE rates. To explain these rates with the maximum suppression of $^7Be$ neutrinos, we need to assume a 
$\langle P_{\nu_e \nu_e} \rangle_{pp}~\geq~0.60$. In this case the fits are obtained, within $3\sigma$ error, for any values of $\sin^2(2\phi)$. See table 4 for more details. 

So a better agreement for the five experiments is found, with 99\% C.L., for 
$\sin^2(2\phi)~\geq~0.50$. The allowed regions that fit the observed rates of the five experiments with 99\% C.L. are shown in the figure 5. In this figure are plotted the calculated rates using the maximum suppression of $^7Be$ neutrinos and $\langle P_{\nu_e \nu_e} \rangle_{pp}~\geq~0.60$. From the figure we see that the allowed regions of 
$\delta m_{21}^2$~x~$\sin^2(2\omega)$ grows when $\sin^2(2\phi)$ rises. For $\sin^2(2\phi)~=~0$,
the range of the parameters are, $0.61~\leq~\sin^2(2\omega)~\leq~1$ and  
$\delta m_{21}^2 \geq 5.0$ . $10^{-11} eV^2$; for $\sin^2(2\phi)~=~0.50$, $0.48~\leq~\sin^2(2\omega)~\leq~1$ and $\delta m_{21}^2 \geq 4.8$ . $10^{-11} eV^2$ and for
$\sin^2(2\phi)~=~1$, $0.29~\leq~\sin^2(2\omega)~\leq~1$ and  
$\delta m_{21}^2 \geq 3.1$ . $10^{-11} eV^2$. The $\chi^2$ analysis for all plots in figure 5 are shown in table 5 and we can observe again that the inclusion of the recoil-electron spectrum gives allways smaller values of $\chi^2$.

\begin{table}
\caption{Best fit points for the $\delta m_{21}^2$ x $\sin^2(2\omega)$ plots shown on figure 5. Column~1 is for the seasonal effect with the rates only and column~2 is with the inclusion of the recoil-electron spectrum.}
\begin{center}
\begin{tabular}{|l|l|l|l|} \hline
$\sin^2(2\phi)$ & $\chi^2_{min}$ & $\sin^2(2\omega)$ & $\delta m_{21}^2 (eV^2)$ \\ \hline
\multicolumn{4}{|c|}{Column 1} \\ \hline
0    & $8.89$ & $0.62$ & $8.32$ . $10^{-11}$ \\ 
0.25 & $9.64$ & $0.96$ & $7.60$ . $10^{-10}$ \\ 
0.50 & $10.5$ & $0.86$ & $7.48$ . $10^{-10}$ \\ 
0.75 & $11.5$ & $0.91$ & $3.47$ . $10^{-10}$ \\ 
1    & $13.9$ & $0.72$ & $3.47$ . $10^{-10}$ \\ \hline
\multicolumn{4}{|c|}{Column 2} \\ \hline
0    & $4.86$ & $0.63$ & $8.90$ . $10^{-11}$ \\ 
0.25 & $5.40$ & $0.56$ & $8.90$ . $10^{-11}$ \\ 
0.50 & $6.13$ & $0.48$ & $8.90$ . $10^{-11}$ \\ 
0.75 & $6.98$ & $0.42$ & $2.75$ . $10^{-10}$ \\ 
1    & $7.56$ & $0.44$ & $4.42$ . $10^{-10}$ \\ \hline
\end{tabular}
\end{center}
\end{table}

\section{Conclusions}

\noindent

The recent Super-Kamiokande data on solar neutrinos brought new constraints on the previously allowed solutions and have restricted them. In this work we show that the vacuum oscillations still give a good explanation to the solar neutrino problem and that three neutrino solutions gives better results. In three neutrino scenario the allowed regions for $\delta m_{21}^2$ and 
$\sin^2(2\omega)$ parameters that fits all experimental data increase as $\sin^2(2\phi)$ grows. 

The calculated suppression rates including the recoil-electron spectrum leads to a reduction of approximately 30\% in the mass regions, but it does not affect the fitting of the data because the averaged probabilities are almost the same as those obtained considering the rates only. An analysis with and without the recoil-electron spectrum shows that the inclusion of this spectrum 
checks better the experimental data. 

A better fit for the experiments are obtained, with 99\% C.L., if we consider the maximum suppression of $^7Be$ neutrinos and a suppression of $pp$ neutrinos from 0 to 40\% as well a suppression of about 60\% of $^8B$ neutrinos. In this case the best agreement with the experimental data is obtained for $\sin^2(2\phi) \geq 0.50$. The inclusion of the recoil-electron spectrum gives a better $\chi^2$ fit than the calculated with rates only, for any value of 
$\sin^2(2\phi)$.  

\section*{Acknowledgments} 

\noindent

We would like to thank the CNPq (Brazilian Council of Scientific and Technologic Developments) for the financial support and the CBPF (Centro Brasileiro de Pesquisas F\'{\i}sicas) for the facilities.

\newpage

\listoffigures

\vspace{.2 true cm}

\noindent
Figure 1 - Normalized spectral sensitivity as a function of the neutrino energy, for the five experiments.

\vspace{.2 true cm}

\noindent
Figure 2 - Limits of variation for $P_{\nu_e \nu_e}$, $P_{\nu_\mu \nu_e}$ and $P_{\nu_\tau \nu_e}$ as a function of $\sin^2(2\phi)$.

\vspace{.2 true cm}

\noindent
Figure 3 - Mass regions versus mixing angles for the BP98 SSM with 99\% C.L. in vacuum, with seasonal effect, for different values of $\sin^2(2\phi)$. The first column is for the rates only and the second column is for the rates with recoil-electron spectrum. The black dot in each plot represents the best fit. 

\vspace{.2 true cm}

\noindent
Figure 4 - Electron neutrino survival probability as a function of the energy for the BP98 SSM with 99\% C.L. in vacuum, for different values of $\sin^2(2\phi)$. The full lines represent the best fit for $P_{\nu_e \nu_e}$ calculated with the rates only (column 1 on figure 3) and the dashed lines the same fits with rates and recoil-electron spectrum (column 2 on figure 3). 

\vspace{.2 true cm}

\noindent
Figure 5 - Mass regions versus mixing angles for the BP98 SSM with 99\% C.L. in vacuum, with seasonal effect, for different values of $\sin^2(2\phi)$ considering maximal suppression of $^7Be$ neutrinos. The first column is for the rates only and the second column is for the rates with recoil-electron spectrum. The black dot in each plot represents the best fit. 


\newpage

\begin{thebibliography}{100}

\bibitem{Bahcall-1} J.N. Bahcall, {\it Neutrino Astrophysics}, 1st edn. (Cambridge University Press, Cambridge 1989)

\bibitem{Pontecorvo} V.N. Gribov and B.M. Pontecorvo, Phys. Lett. B {\bf 28}, 493 (1969)

\bibitem{MSW} S.P. Mikheyev and A. Yu. Smirnov, Sov. J. Nucl. Phys. {\bf 42}, 913 (1986); Nuovo Cimento C {\bf 9}, 17 (1986); L. Wolfenstein, Phys. Rev. D {\bf 17}, 2369 (1978)

\bibitem{Borexino} R.S. Raghavan, Science {\bf 267}, 45 (1995); C. Arpesella {\it et al.}, {\it Borexino Proposal}, vols. 1 and 2, 1st edn. (G. Bellini {\it et al.}, University of Milano, Milano, 1992)

\bibitem{Homestake} B. Cleveland {\it et al.}, Astrop. Journal {\bf 496}, 505 (1998); B. Cleveland {\it et al.}, Nucl. Phys. B (Proc. Suppl.) {\bf 38}, 47 (1995); R. Davis, Prog. Part. Nucl. Phys. {\bf 32}, 13 (1994)

\bibitem{GALLEX} GALLEX Collaboration, P. Anselmann {\it et al.}, Phys. Lett. B {\bf 342}, 440 (1995); GALLEX Collaboration, W. Hampel {\it et al.}, Phys. Lett. B {\bf 388}, 364 (1996)

\bibitem{SAGE} SAGE Collaboration, V. Gavrin {\it et al.}, {\it Neutrino 96, Proc. of the XVII Int. Conf. on Neutrino Physics and Astrophysics, Helsinki, 1997}, edited by World Scientific (Singapore, 1997)

\bibitem{Kamiokande} Kamiokande Collaboration, Y. Fukuda {\it et al.}, Phys. Rev. Lett. {\bf 77}, 1683 (1996)

\bibitem{Super-K} Super-Kamiokande Collaboration, Y. Fukuda {\it et al.}, Phys. Rev. Lett. {\bf 81}, 1158 (1998)

\bibitem{BP98} J.N. Bahcall {\it et al.}, Phys. Lett. B {\bf 433}, 1 (1998); Phys. Rev. D {\bf 58}, 096016 (1998)

\bibitem{Bahcall-87} J.N. Bahcall, Rev. of Mod. Physics {\bf 59}, 505 (1987); J.N. Bahcall {\it et al.}, Phys. Rev. C {\bf 55}, 494 (1997); J.N. Bahcall {\it et al.}, Phys. Rev. D {\bf 51},  6146 (1995); B. Fa\"id {\it et al.}, Phys. Rev. D {\bf 55}, 1353 (1997); B. Fa\"id {\it et al.}, Astrop. Phys. {\bf 10}, 93 (1999)

\bibitem{Fogli} G.L. Fogli {\it et al.}, Phys. Rev. D {\bf 49}, 3626 (1994)

\bibitem{Barger} V. Barger {\it et al.}, Phys. Rev. D {\bf 43}, 1110 (1991)

\bibitem{Krastev} P.I. Krastev and S.T. Petcov, CERN-preprint, CERN-TH.6401/92, 1992 

\bibitem{Calabresu} E. Calabresu {\it et al.}, Astrop. Phys. {\bf 4}, 159 (1995)

\end{thebibliography}
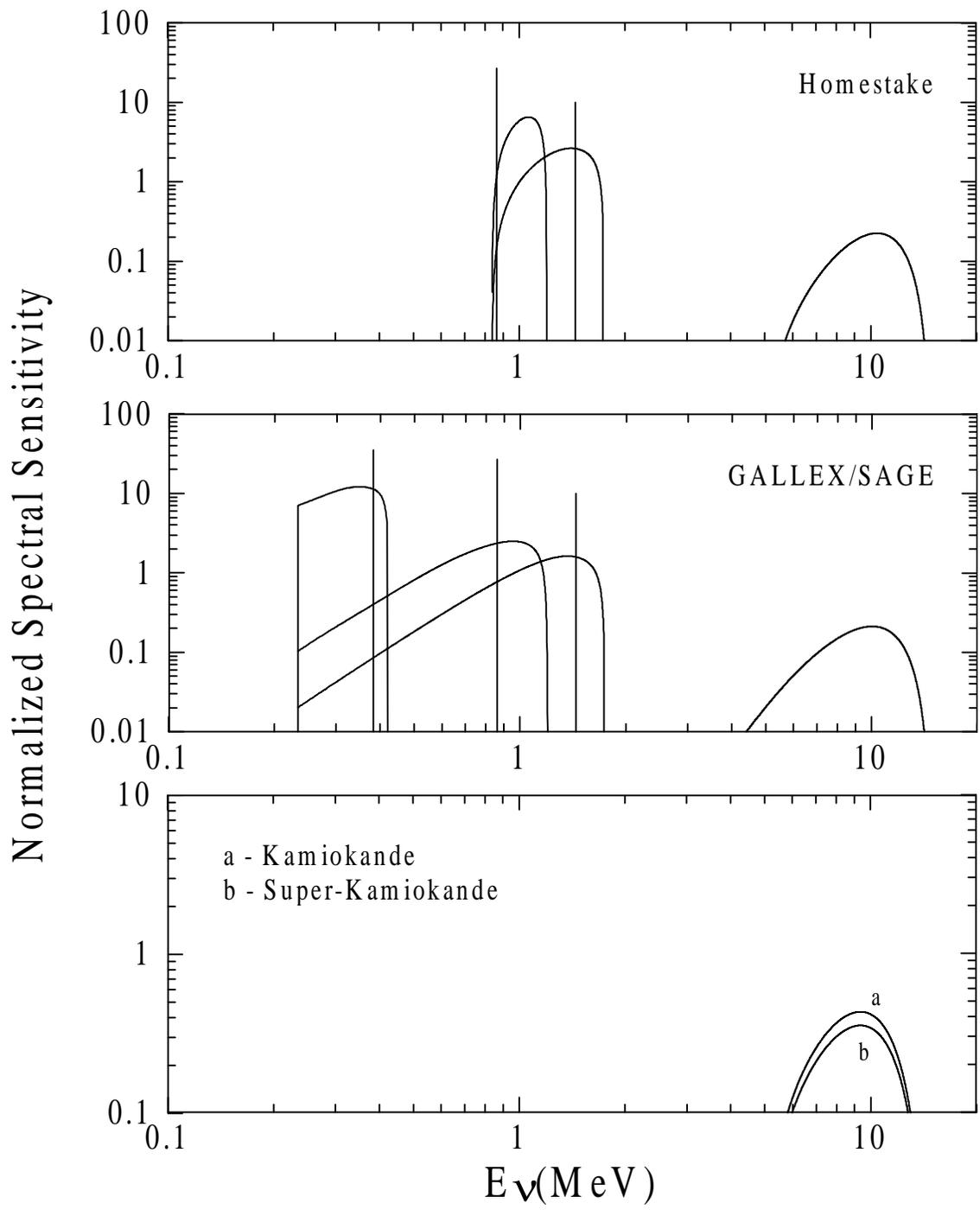
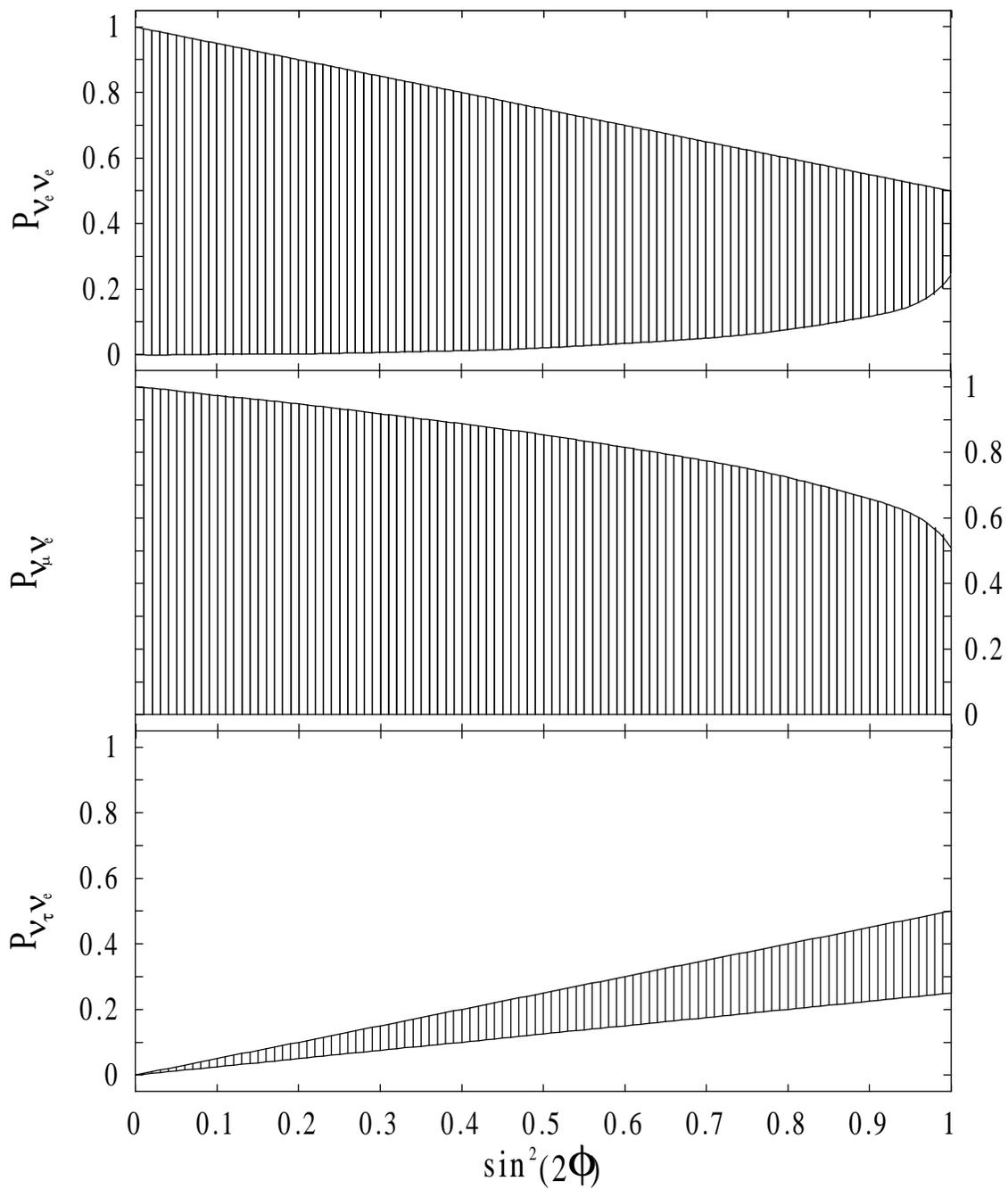
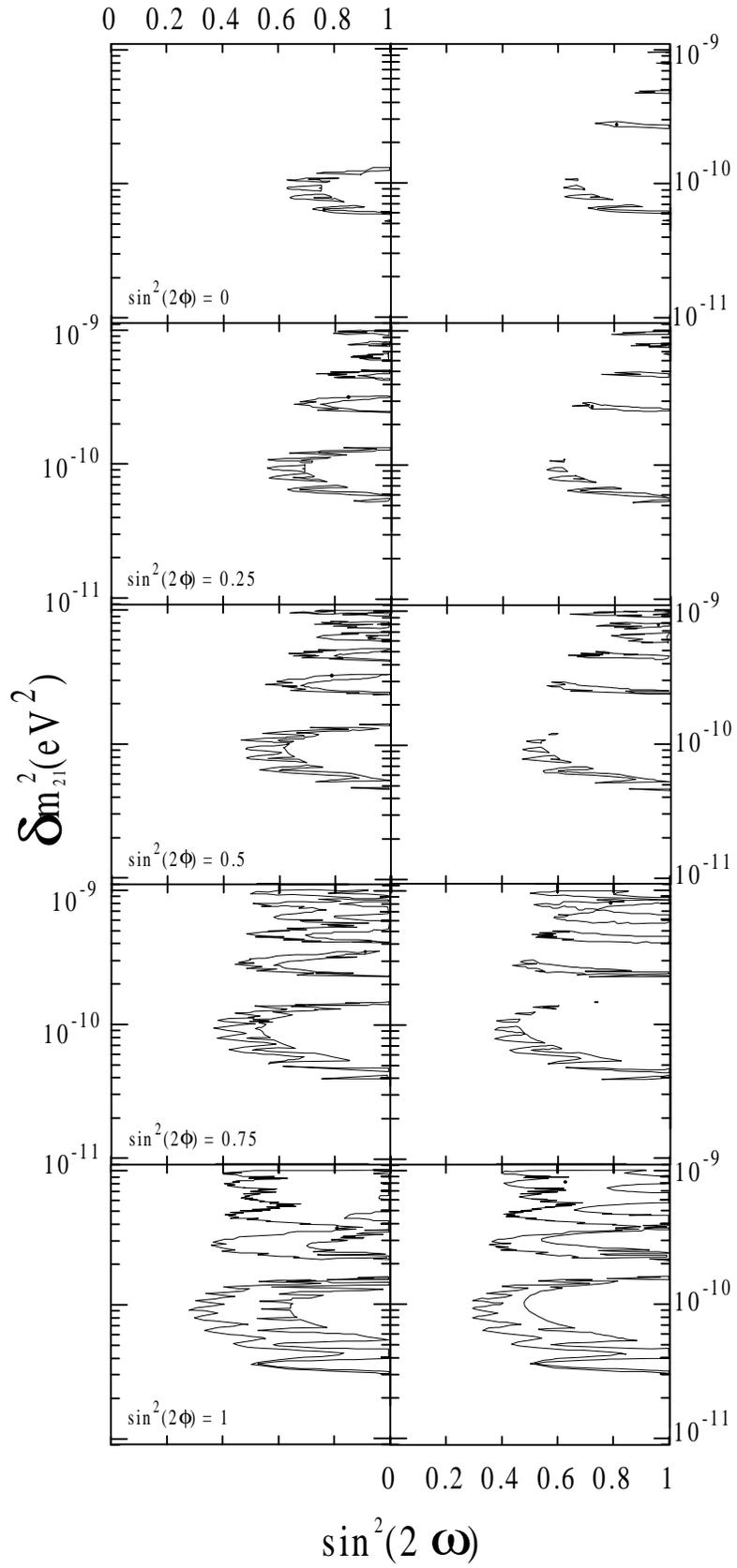
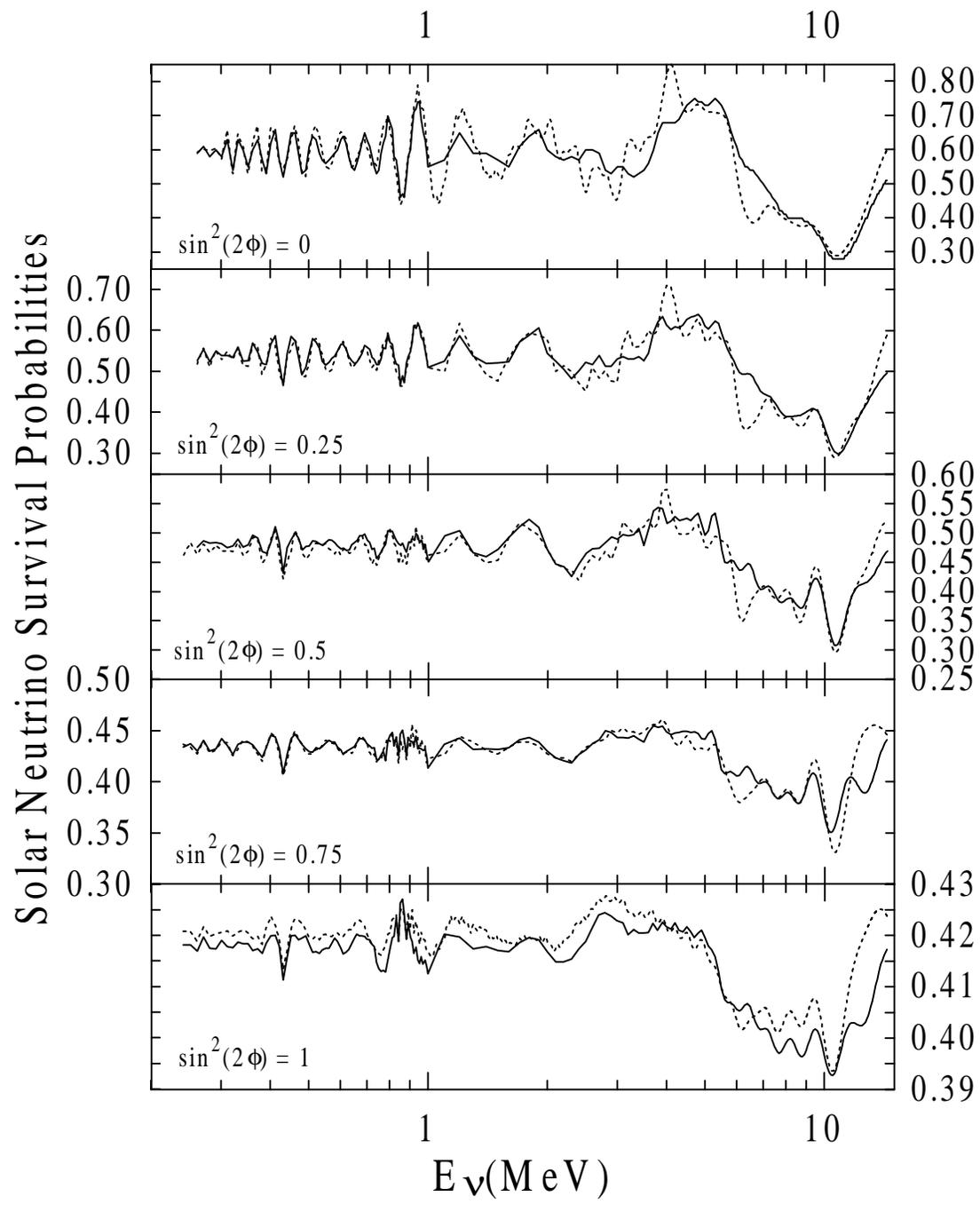
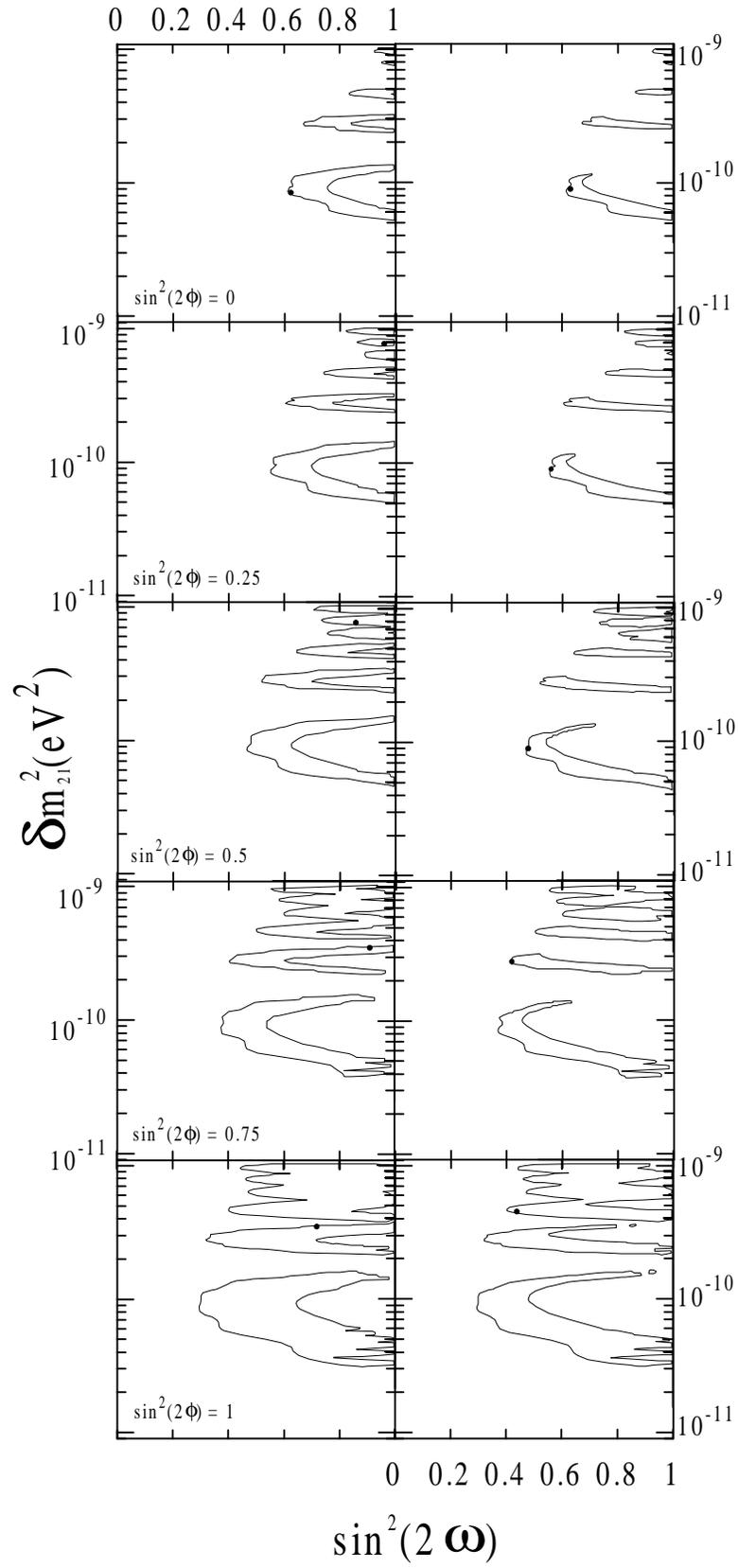
\end{document}